\documentclass[aps,prl,twocolumn,amsmath,showpacs,showkeys,superscriptaddress,floatfix]{revtex4}

\usepackage{amsmath}
\usepackage{epsfig}
\usepackage{graphicx}
\usepackage{graphics}
\usepackage{url}
\usepackage[english]{babel}
\usepackage{dcolumn}
\usepackage{color}
\usepackage[squaren]{SIunits}
\begin{document}
\title{Direct optical state preparation of the dark exciton in a quantum dot}

\author{S.~L\"uker}
\affiliation{Institut f\"ur Festk\"orpertheorie, Universit\"at M\"unster,
Wilhelm-Klemm-Str.~10, 48149 M\"unster, Germany}

\author{T.~Kuhn}
\affiliation{Institut f\"ur Festk\"orpertheorie, Universit\"at M\"unster,
Wilhelm-Klemm-Str.~10, 48149 M\"unster, Germany}

\author{D.~E.~Reiter}
\affiliation{Institut f\"ur Festk\"orpertheorie, Universit\"at M\"unster,
Wilhelm-Klemm-Str.~10, 48149 M\"unster, Germany}
\affiliation{Department of Physics, Imperial College London, South Kensington Campus,
London SW7 2AZ, United Kingdom}

\date{\today}

\begin{abstract}
Because of their weak coupling to the electromagnetic field dark excitons in
semiconductor quantum dots possess extremely long lifetimes, which makes them
attractive candidates for quantum information processing. On the other hand,
preparation and manipulation of dark states is challenging, because commonly
used optical excitation mechanisms are not applicable. We propose a new,
efficient mechanism for the deterministic preparation of the dark exciton
exploiting the application of a tilted magnetic field and the optical
excitation with a chirped, i.e., frequency modulated, laser pulse.
\end{abstract}



\maketitle

The optical properties of an exciton in a semiconductor quantum dot (QD) are
determined by the combination of spin states of electron and hole. In the
most common case of heavy hole excitons among the four different states the
two excitons with antiparallel electron and hole spin are optically active or
bright. The other two excitons with parallel spin do not couple to the light
field and are called dark. The bright states have been proposed for various
applications, e.g., as single photon sources
\cite{yuan2002ele,santori2001tri,press2007pho} or qubits
\cite{bonadeo1998coh,biolatti2000qua,troiani2000exp,damico2002all,boyle2008two,michaelis2010coh},
and a large amount of research has been dedicated to their optical state
preparation and control \cite{ramsay2010are,reiter2014the}. Dark excitons
seem to be optically inaccessible and have almost no influence on bright
excitons; therefore they are often neglected. But dark excitons offer
unexplored potential for applications as photon memory or for quantum
operations \cite{korkusinski2013ato,poem2010acc}, because their lifetime
exceeds the one of bright excitons by far
\cite{poem2010acc,mcfarlane2009gig}. For this reason it is very attractive to
use dark excitons for quantum information applications and to overcome the
huge challenge to prepare dark excitons in a controlled and direct manner.

In this Communication we propose a method to optically prepare dark excitons
in a deterministic way. Other methods to optically access dark excitons have
been proposed recently, either relying on the relaxation from higher excited
and biexciton states
\cite{poem2010acc,korkusinski2013ato,smolenski2015mec,schmidgall2015all} or
on valence band mixing caused by asymmetry \cite{schwartz2015det}. In
contrast, in our proposal dark excitons are directly excited from the QD
ground state and the excitation mechanism can be applied to all typical II-VI
and III-V QDs. Our preparation scheme uses two simple ingredients: a chirped
laser pulse and a tilted magnetic field. Chirped laser pulses have already
shown great potential for preparation of bright excitons and biexcitons via
adiabatic rapid passage (ARP)
\cite{wu2011pop,simon2011rob,mathew2014sub,luker2012inf,gawarecki2012dep,glassl2013bie,debnath2012chi,malinovsky2001gen}.
The in-plane component of a magnetic field couples bright and dark exciton
via a spin-flip \cite{bayer2000spe,besombes2000fin,kazimierczuk2011mag} and
therefore transfers oscillator strength to the dark exciton. We will show how
the clever combination of these two ingredients allows for an efficient dark
exciton preparation.

We consider a self-assembled QD, where the uppermost valence band is of
heavy-hole type, i.e., the holes have a spin of $S^h_z = \pm 3/2$ while the
electrons have a spin of $S^e_z = \pm 1/2$. This leads to the formation of
four single exciton states with total spin $J_z=S^h_z+S^e_z \in\{\pm1,\pm
2\}$. According to the dipole selection rules only excitons with a total spin
of $+1$ ($-1$) can be optically excited, which corresponds to excitation with
positive (negative) circularly polarized laser pulses. We label the
corresponding bright states $|b^+\rangle$ ($|b^-\rangle$). Excitons with
total spin of $+2$ ($-2$) are not optically active and labeled $|d^+\rangle$
($|d^-\rangle$). The biexciton state is denoted by $|B\rangle$. The short
range exchange interaction gives rise to a splitting $\delta_0$ between
bright and dark excitons. In most real QDs the bright excitons
$|b^\pm\rangle$ are coupled by the long range exchange interaction leading to
linearly polarized excitons with a splitting $\delta_1$. Analogously, the
dark excitons $|d^\pm\rangle$ are coupled and split by $\delta_2$.

\begin{figure}[t]
\includegraphics[width=\columnwidth]{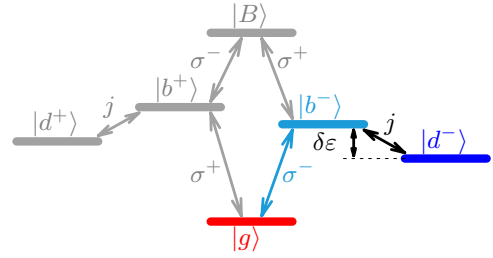}
\caption{(Color online) Schematic level diagram.}\label{fig1}
\end{figure}

An external magnetic field $\mathbf{B}$ is used to control the level
structure of the system and the coupling between the states. The out-of-plane
component of the magnetic field $B_z$ provides a control parameter for the
exciton energies via the Zeeman shifts of the electron (hole) determined by
the out-of-plane $g$-factors $g_{e,z}$ ($g_{h,z}$) \cite{bayer1999ele}. The
in-plane magnetic field $B_x$ induces spin flips of the electron (hole) via
the in-plane $g$-factors $g_{e,x}$ ($g_{h,x}$). While the model is applicable
to many materials, here, we take GaAs material parameters
\cite{bayer2000spe,bayer2002fin,kuther1998zee}. The Hamiltonian together with
the material parameters is given in the supplemental information.

The choice of the magnetic field and in particular the ratio between
out-of-plane and in-plane component of the magnetic field is crucial in our
set-up. On the one hand the in-plane magnetic field is needed to couple
bright and dark exciton to enable dark exciton state preparation, on the
other hand it should be weak such that bright and dark exciton are still
well-defined. We find that these conditions are fulfilled for $B_z =
4\,{\rm{T}}$  and $B_x = 3\,{\rm{T}}$. The level structure of the six states
for these parameters is depicted in Fig~\ref{fig1}.

It turns out that for the chosen magnetic field, when restricting ourselves
to the case of excitation with $\sigma^-$-polarized light, the system can be
reduced to a three-level model consisting of the states $|g\rangle$,
$|b^-\rangle$, and $|d^-\rangle$ (colored states in Fig.~\ref{fig1}).
Therefore, in the following discussions we will concentrate on this reduced
model and omit the polarization superscript ``$-$'' of the exciton states.
The three-level model is characterized by the energy difference $\delta
\epsilon = \delta_0 - g_{e,z} \mu_B B_z = 0.43\,{\rm{meV}}$ and the coupling
$j = \frac{1}{2}\mu_B g_{e,x} B_x= 0.056\,{\rm{meV}}$. With these values, the
admixture of the bright exciton to the dark exciton is only about 1.5\,\%.
Thus the lifetime of the dark exciton is expected to be still about two
orders of magnitude larger than the lifetime of the bright exciton
\cite{stevenson2004tim}.

\begin{figure}[t]
\includegraphics[width=\columnwidth]{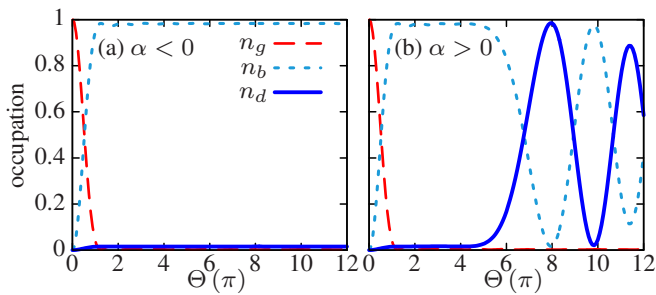}
\caption{(Color online) Final occupation of ground state $n_g$
(red dashed line), bright exciton $n_b$ (light blue dotted
line), and dark exciton $n_d$ (solid blue line) after
excitation with a pulse with pulse area $\Theta$ and for (a) negative chirp $\alpha =
-40\,{\rm{ps}^2}$, and (b) positive chirp $\alpha =
+40\,{\rm{ps}^2}$.}\label{fig2}
\end{figure}

We consider an excitation with a chirped laser pulse, which has been shown to
yield a population inversion of the bright exciton system, being stable
against small changes of the pulse parameters
\cite{wu2011pop,simon2011rob,reiter2014the,debnath2012chi}. The instantaneous
frequency of the pulse $\omega(t) = \omega_0 + at$ changes with the chirp
rate $a$, while its envelope is characterized by the pulse area $\Theta$ and
the duration $\tau$. Such a linearly chirped Gaussian pulse can be obtained
by sending a transform-limited Gaussian laser pulse with duration $\tau_0$
through a chirp filter with chirp coefficient $\alpha$. For our calculations
we use pulses with an initial duration $\tau_0 = 3.5\,{\rm{ps}}$ and a chirp
coefficient $\alpha = \pm 40\,{\rm{ps}}^2$, which yields chirped pulses with
$\tau \approx 12\,{\rm{ps}}$ and $a \approx \pm 0.023\,{\rm{ps}}^{-2}$ (for
details see supplemental information). The central frequency $\omega_0$ at
the pulse maximum is taken to be in resonance with the bright exciton
transition.

\begin{figure}[t]
\includegraphics[width=\columnwidth]{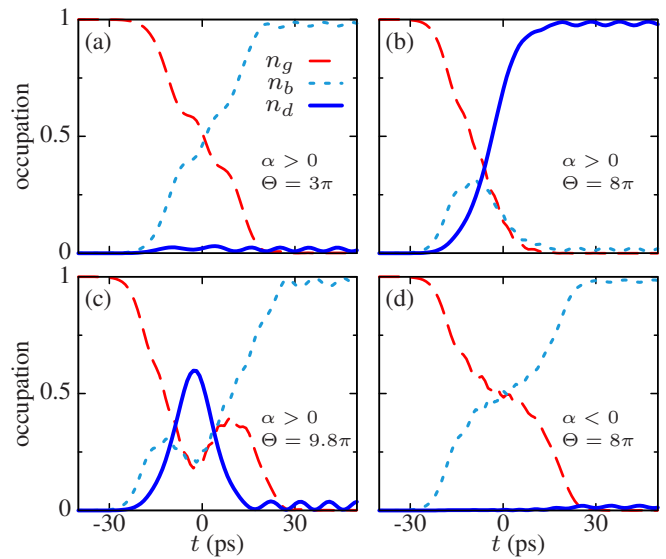}
\caption{(Color online) Temporal evolution of the occupation of
ground state $n_g$ (red dashed line), bright exciton
$n_b$ (light blue dotted line), and dark exciton $n_d$ (solid
blue line) for $\alpha = 40\,{\rm{ps}}^2 $ and (a) $\Theta =
3\pi$, (b) $\Theta = 8\pi$, and (c) $\Theta = 9.8\pi$. (d)
Same as part (b), but for negative chirp $\alpha =
-40\,{\rm{ps}}^2$.}\label{fig3}
\end{figure}

For state preparation the common quantity to describe the excitation fidelity
is the final occupation after the pulse \cite{ramsay2010are, reiter2014the}.
Typically the final occupations are shown as a function of the laser pulse
power expressed in terms of the pulse area $\Theta$. Figure~\ref{fig2}(a)
shows the final occupation of the ground state $n_g$, the bright exciton
$n_b$, and the dark exciton $n_d$ as a function of pulse area for a
negatively chirped pulse with $\alpha = -40\,{\rm{ps}}^2$. We find that the
behavior is equivalent to ARP in a two-level system. As soon as the pulse
area exceeds the adiabatic threshold a robust occupation of the bright
exciton is seen. The dark exciton is completely unaffected and its occupation
remains at $n_d = 0$.

The situation changes dramatically when the sign of the chirp is reversed,
i.e., for excitation with a positively chirped pulse, as depicted in
Fig~\ref{fig2}(b). While for pulse areas up to $\Theta \approx 5\pi$ the
behavior is similar to the two-level case, for higher pulse areas the dark
exciton becomes significantly occupied. At $\Theta = 8\pi$ a maximal
occupation of approximately $n_d = 1$ is achieved. This is a remarkable
result, since the dark exciton is not directly affected by the chirped laser
pulse, but only coupled indirectly via the bright exciton to the light field.
When the pulse area is increased further, we find that the final occupation
oscillates between bright and dark exciton. To validate the approximation of
the three-level system, we performed calculations in the full six-level
system which showed that the sum of the occupations of the neglected states
remains indeed well below 1\%.

To understand the dynamics during the excitation in Fig.~\ref{fig3} we have
plotted the temporal evolution of the state occupations for different
excitation conditions. Figure~\ref{fig3}(a) shows the case of a positively
chirped laser pulse ($\alpha = 40\,{\rm{ps}}^2$) and a pulse area of $\Theta
= 3\pi$. During the excitation the occupation of the bright exciton increases
monotonically to $n_b = 1$, while the ground state occupation goes to zero.
The dark exciton is nearly unaffected. Hence the complete system can be
considered as an effective two-level system consisting of ground state and
bright exciton. After the pulse there are small oscillations between bright
and dark exciton caused by the in-plane magnetic field. In agreement with our
choice of magnetic field, these oscillations are so small that the
discrimination between bright and dark exciton is still valid. To account for
these oscillations all final occupations shown in this Communication are averaged
over one period.

Figure~\ref{fig3}(b) shows the dynamics at $\Theta = 8\pi$, where the first
maximum of the dark exciton occupation occurs. Initially the ground state
occupation decreases, while the occupation of the bright exciton increases.
Then, also the dark exciton becomes increasingly populated. The bright
exciton occupation reaches a maximum around $t = -10\,{\rm{ps}}$ and
subsequently drops to $n_b \approx 0$, just as the ground state occupation.
The occupation of the dark exciton increases monotonically up to $n_d \approx
1$. The dynamics at $\Theta=9.8\pi$, i.e., at the first minimum of the final
dark exciton occupation, is depicted in Fig.~\ref{fig3}(c). First the
population of the ground state drops down in favor of the bright and dark
exciton occupation, until all three states are almost equally populated.
During the peak intensity of the laser pulse all three occupations oscillate,
the dark exciton occupation oscillating opposite to the ground state and
bright exciton occupations. After one oscillation the dark exciton occupation
returns to $n_d \approx 0$, while the bright exciton becomes almost
completely populated. To complete the discussion of the dynamics,
Fig.~\ref{fig3}(d) shows the occupations for negative chirp $\alpha =
-40\,{\rm{ps}}^2$ and $\Theta = 8\pi$. We find that a population inversion
between ground state and bright exciton takes place, but even for such high
pulse areas the dark exciton occupation remains at $n_d \approx 0$. This
confirms that the system can be reduced to a two-level system for excitation
with negatively chirped laser pulses.

\begin{figure}[ht]
\includegraphics[width=\columnwidth]{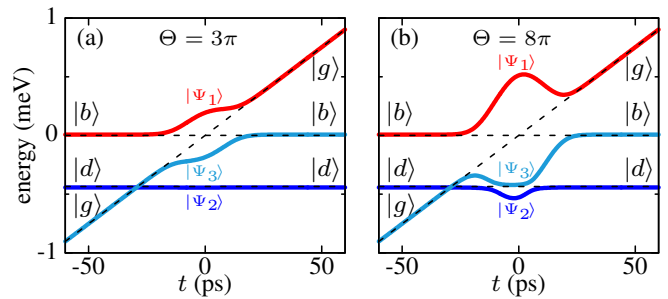}
\caption{(Color online) Instantaneous eigenenergies corresponding
to the dressed states $|\Psi_1\rangle$, $|\Psi_2\rangle$, and
$|\Psi_3\rangle$, for (a) $\Theta = 3\pi$ and (b) $\Theta = 8\pi$. Dashed lines
illustrate the evolution of the states
without interaction.}\label{fig4}
\end{figure}

A well established picture to understand the dynamics of a system under
strong excitation is the dressed state picture \cite{tannor2007int}. Dressed
states are the instantaneous eigenstates of the system, which are obtained by
a diagonalization of the Hamiltonian, including light-matter interaction and
magnetic field coupling. In general, each dressed state is composed of a
mixture of all bare states $|g\rangle$, $|b\rangle$, and $|d\rangle$. Only
for vanishing interaction a dressed state is identical to a bare state. In
our system this happens in the limit of times long before and after the pulse
because then also the coupling by $B_x$ can be neglected since $\delta
\epsilon \gg j$.

Figure~\ref{fig4} shows the evolution of the instantaneous eigenenergies
corresponding to the indicated dressed states in a frame rotating with the
laser frequency. In Fig.~\ref{fig4}(a) the pulse area is $\Theta = 3\pi$,
which is in the parameter region where we found a behavior like in a
two-level system. The system is initially in the ground state $|g\rangle$,
which agrees with the dressed state $|\Psi_3\rangle$ for times long before
the pulse. The system evolves adiabatically along the branch which is related to $|\Psi_3\rangle$ in Fig.~\ref{fig4}(a). Around $t = -30\,{\rm{ps}}$ this branch crosses the
dressed state $|\Psi_2\rangle$, which corresponds to the dark exciton. The
system evolves straight through this crossing, because the laser pulse is not
yet effective and there is no direct coupling between ground state and dark
exciton. Due to the chirp the system traverses afterwards a broad
anti-crossing with the state $|\Psi_1\rangle$. This state agrees with the
bright exciton for times long before the pulse. Because of the chirp the
character of the dressed states changes during the pulse. For times long
after the pulse the state $|\Psi_3\rangle$ can be identified as the bright
exciton $|b\rangle$, which gives rise to the ARP effect \cite{tannor2007int}.

Let us now consider the evolution of the eigenstates at $\Theta = 8\pi$,
where the dark exciton becomes populated. Analogous to the case of low pulse
area the system starts on the lowest branch $|\Psi_3\rangle$ and passes the
crossing with $|\Psi_2\rangle$. The main difference is the size of the
splitting between $|\Psi_1\rangle$ and $|\Psi_3\rangle$ around $t = 0$, which
is now much larger because of the high pulse intensity. This results in a
strong bending of $|\Psi_3\rangle$ such that it approaches $|\Psi_2\rangle$.
Because $|\Psi_3\rangle$ and $|\Psi_2\rangle$ are coupled, in accordance to
the coupling between bright and dark state induced by the in-plane magnetic
field, two additional anti-crossings between the two branches emerge.
However, the splitting of these anti-crossings is small, such that the system
cannot evolve adiabatically through them anymore. Instead, transitions
between the involved branches occur resulting in quantum beats between the
two dressed states. This is confirmed by our finding in Fig.~\ref{fig3}(c),
where the occupations of the states oscillate in time and also by the
oscillatory behavior shown in the final occupation in Fig.~\ref{fig2}(b). The
final state of the oscillation is determined by the laser pulse intensity,
i.e., the mixing of the states, and the duration of the quantum beats.
Accordingly, after the second anti-crossing the system can end up either in
$|\Psi_2\rangle$ or $|\Psi_3\rangle$, or any superposition between them. A
similar picture is given by the optical Stark effect, by which for strong
pulses bright and dark exciton can be brought into resonance such that
quantum beats occur \cite{reiter2012spi,reiter2013opt}.

We again complete the discussion considering negative chirps. The change of
the sign of the chirp is equivalent to a time reversal. This means that the
dressed state diagrams have to be read backwards. When the QD is initially in
the ground state the system now evolves along the upper branch
$|\Psi_1\rangle$. Because $|\Psi_1\rangle$ is the uppermost dressed state and
this state is always well separated from the other states, no transitions to
$|\Psi_2\rangle$ are possible, i.e., a population of the dark exciton cannot
be achieved and the system evolves adiabatically into the bright exciton
state. However, if the QD has been driven into the dark exciton state by a
pulse with positive chirp, the evolution caused by this pulse is reversed and
the system evolves back into the ground state. Thus by applying a sequence of
pulses with alternately positive and negative chirp the QD can be switched
back and forth between ground state and dark exciton.

\begin{figure}[t]
\includegraphics[width=\columnwidth]{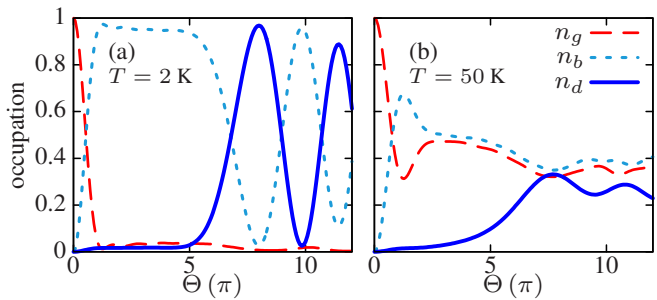}
\caption{(Color online) Same as Fig.~\ref{fig2}(b) including electron-phonon
interaction at (a) $T = 2\,$K and (b) $T = 50\,$K.}\label{fig5}
\end{figure}

A realistic QD is not completely isolated but interacts with its environment.
Previous studies have shown that the coupling to phonon modes of the
surrounding material is the main source of decoherence in QDs
\cite{ramsay2010are,reiter2014the}. In the following we consider the coupling
to longitudinal acoustic (LA) phonons via the deformation potential using
GaAs material parameters within a well-established fourth-order correlation
expansion (cf. supplemental information) \cite{krugel2005the,luker2012inf}.
Figure~\ref{fig5} shows the influence of the phonons on the final occupation
of the three states for excitation with positively chirped laser pulses at
two different temperatures $T=2$~K and $T=50$~K.

Previous calculations in the two-level system have shown that the effect of
phonons on state preparation using ARP can be described by transitions
between the dressed states \cite{luker2012inf,gawarecki2012dep}. For low
temperatures, only phonon emission takes place, while phonon absorption
processes are suppressed. Therefore, we do not expect a significant influence
of the phonons, when positively chirped pulses are used for the dark exciton
preparation, because the evolution takes place on the lowest branch of the
dressed states. Even when the system is on the central branch after the
anti-crossing no phonon-assisted transitions occur to the lower branch
$|\Psi_2\rangle$, because the phonon coupling strength is the same for both
bright and dark exciton. Indeed, as seen in Fig.~\ref{fig5}(a), the influence
of phonons is almost negligible for $T = 2$~K and the behavior is similar to
the phonon free case shown in Fig.~\ref{fig2}(b). For higher temperatures,
phonon absorption processes become possible, which results in an almost equal
occupation of the three states as shown in Fig.~\ref{fig5}(b), which is
similar to the behavior of ARP in a two-level system \cite{luker2012inf}.

%%%%%%%%%%%%%%%%%CONCLUSIONS%%%%%%%%%%%%%%%%%%%%%%%%%%%%%%%%%%%%

In conclusion, we have  shown that the excitation with a chirped laser pulse
in combination with an external magnetic field provides a powerful tool for
the optical preparation of dark excitons in QDs. The direct and deterministic
preparation of the dark exciton is a crucial step towards the usage of dark
excitons for applications, e.g., in quantum information processing.

\begin{acknowledgements}
The authors thank P. Kossacki and M. Goryca for valuable discussions.
Financial support from the DAAD within the P.R.I.M.E. programme is gratefully
acknowledged.
\end{acknowledgements}

%\bibliography{std,qds,review_books,review_phonon,mn}

\end{document}

\title{Supplemental material: Direct optical state preparation of the dark exciton in a quantum dot}

\author{S.~L\"uker}
\affiliation{Institut f\"ur Festk\"orpertheorie, Universit\"at M\"unster,
Wilhelm-Klemm-Str.~10, 48149 M\"unster, Germany}

\author{T.~Kuhn}
\affiliation{Institut f\"ur Festk\"orpertheorie, Universit\"at M\"unster,
Wilhelm-Klemm-Str.~10, 48149 M\"unster, Germany}

\author{D.~E.~Reiter}
\affiliation{Institut f\"ur Festk\"orpertheorie, Universit\"at M\"unster,
Wilhelm-Klemm-Str.~10, 48149 M\"unster, Germany}
\affiliation{Department of Physics, Imperial College London, South Kensington Campus,
London SW7 2AZ, United Kingdom}

\maketitle

We present the Hamiltonian of the full six-level system together with the
chosen material parameters. Furthermore, we introduce the reduced three-level
system, which has been used for most of the calculations.

\begin{section}{Six-level system}
The six-level system consists of the ground state $|g\rangle$ and all
excitonic states that can be formed from the uppermost valence band state,
taken to be of heavy hole type, and the lowest conduction band state. These
are the bright excitons $|b^+\rangle$ ($|b^- \rangle$) with angular momenta
$+1$ ($-1$), the corresponding dark excitons $|d^+\rangle$ ($|d^-\rangle$)
with angular momenta $+2$ ($-2$), and the biexciton $|B\rangle$. The
Hamiltonian for this system consists of four parts
\begin{align*}
H = H_{\textrm{exciton}} + H_{\textrm{exchange}} + H_{\textrm{spin-flip}} + H_{\textrm{carrier-light}},
\end{align*}
where $H_{\textrm{exciton}}$ describes the energetic structure of the exciton
system including energy shifts caused by the short-range exchange interaction
and the magnetic field component in growth direction, while the other three
contributions describe couplings among these states: $H_{\textrm{exchange}}$
represents the long-range electron-hole exchange interaction,
$H_{\textrm{spin-flip}}$ describes the influence of the external in-plane
magnetic field, and $H_{\textrm{carrier-light}}$ denotes the carrier-light
coupling. Let us go through the four parts one by one.

The exciton Hamiltonian is given by
\begin{align*}
H_{\textrm{exciton}} = \, & E_{b^+}|b^+\rangle\langle b^+|
                          + E_{b^-}|b^-\rangle\langle b^-| \\
                          &+   E_{d^+} |d^+\rangle\langle d^+|
                          + E_{d^-}|d^-\rangle\langle d^-|
                          + E_B |B\rangle\langle B|
\end{align*}
with the corresponding energies
\begin{align*}
E_{b^+} & = E_X+\frac{\delta_0}{2}-\frac{\mu_B B_z}{2}\left(3g_{h,z}-g_{e,z}\right), \\
E_{b^-} & = E_X+\frac{\delta_0}{2}+\frac{\mu_B B_z}{2}\left(3g_{h,z}-g_{e,z}\right), \\
E_{d^+} & = E_X-\frac{\delta_0}{2}-\frac{\mu_B B_z}{2}\left(3g_{h,z}+g_{e,z}\right), \\
E_{d^-} & = E_X-\frac{\delta_0}{2}+\frac{\mu_B B_z}{2}\left(3g_{h,z}+g_{e,z}\right),\\
E_B     & = 2E_X-\Delta_B,
\end{align*}
where $E_X$ is the bare exciton energy, $\Delta_B$ the biexciton binding
energy, $\delta_0$ the energy splitting between bright and dark excitons
caused by the short-range exchange interaction, $g_{e,z}$ ($g_{h,z}$) is the
electron (hole) out-of-plane $g$-factor, $\mu_B$ the Bohr magneton, and $B_z$
the out-of plane component of the magnetic field. The energy of the ground
state has been taken to be zero. It is easy to see that the out-of-plane
magnetic field $B_z$ is a control parameter for the energetic ordering of the
exciton states.

The long-range exchange interaction
\begin{align*}
H_{\textrm{exchange}} = & \frac{\delta_1}{2}\left(|b^+\rangle\langle b^-| +
|b^-\rangle\langle b^+|\right) \\
&+\frac{\delta_2}{2}\left(|d^+\rangle\langle d^-| +
|d^-\rangle\langle d^+|\right)
\end{align*}
couples excitons with opposite angular momenta, where $\delta_1$ and
$\delta_2$ are the exchange splittings of bright and dark excitons,
respectively.

The Hamiltonian describing spin-flips caused by an in-plane magnetic field
$B_x$ reads
\begin{align*}
H_{\textrm{spin-flip}} = & -\frac{g_{e,x} \mu_B B_x}{2}
 \left(|b^+\rangle\langle d^+| +
|b^-\rangle\langle d^-| + \textrm{h.c.}\right) \\
& - \frac{g_{h,x} \mu_B B_x}{2} \left(|b^+\rangle\langle d^-| +
|d^+\rangle\langle b^-| + \textrm{h.c.} \right),
\end{align*}
where $g_{e,x}$ ($g_{h,x}$) are the electron (hole) in-plane g-factors and
h.c. denotes the Hermitian conjugate.

The carrier-light coupling is modeled in the usual rotating wave and dipole approximation via
\begin{align*}
H_{\textrm{carrier-light}} = &
 - \frac{\hbar}{2} \left[ \Omega^{+}(t)\left(
|b^{+}\rangle\langle g| + |B\rangle\langle b^{-}|\right) + \textrm{h.c.} \right] \nonumber \\
& - \frac{\hbar}{2} \left[ \Omega^{-}(t)\left(
|b^{-}\rangle\langle g| + |B\rangle\langle b^{+}|\right) + \textrm{h.c.} \right],
\end{align*}
where $\hbar\Omega^{\pm}(t) =  2 M_{0} E^{\pm}(t)$ with $E^{\pm}(t)$ being
the positive frequency component of the electric field of the $\sigma^\pm$
circularly polarized light field, and $M_0$ is the dipole matrix element.

The model is applicable to quantum dots of many materials. To be specific,
here we take values typical for GaAs. The $g$-factors of electron and hole
are $g_{e,z} = -0.8$, $g_{e,x} = -0.65$, $g_{h,z}=-2.2$, and $g_{h,x} =
-0.35$ \cite{bayer2000spe}. The exchange splittings are $\delta_0 =
0.25\,{\rm{meV}}$ , $\delta_1 = 0.12\,{\rm{meV}}$, and $\delta_2 =
0.05\,{\rm{meV}}$ \cite{bayer2002fin}. The biexciton binding energy is
$\Delta_B = 3\,{\rm{meV}}$ \cite{kuther1998zee}.

\end{section}

\begin{section}{Three-level system}

For the considered parameters and negatively circularly polarized light the
system can be reduced to a three-level system by neglecting the positively
polarized excitons and the biexciton. The three-level system then consists of
the ground state $|g\rangle$, the bright exciton $|b\rangle = |b^-\rangle$,
and the dark exciton $|d\rangle = |d^-\rangle$. The Hamiltonian reduces to
\begin{align*}
H = &\, E_b|b\rangle\langle b| + E_d|d\rangle\langle d|
-  \frac{1}{2} g_{e,x}\mu_B B_x\left( |d\rangle\langle b| + |b\rangle\langle d| \right)\\
&- \frac{\hbar}{2} \left[ \Omega(t) |b\rangle\langle g| +
\Omega^{*}(t) |g\rangle\langle b| \right] .
\end{align*}

To analyze the influence of the environment we include the coupling of the
three-level system to longitudinal acoustic phonons via the pure dephasing
mechanism. The Hamiltonian of the carrier-phonon coupling reads
\begin{align*}
H_{\textrm{carrier-phonon}} =\, &  \hbar\sum_{\mathbf{q}}\omega_{\mathbf{q}}b^{\dag}_{\mathbf{q}} b_{\mathbf{q}} \\
& + \hbar\left(|b\rangle\langle b| + |d\rangle\langle d|\right)\sum_{\mathbf{q}}
\left(g_{\mathbf{q}}b_{\mathbf{q}}+g^*_{\mathbf{q}}b^{\dag}_{\mathbf{q}}\right).
\end{align*}
The operators $b^\dag_{\mathbf{q}}$ ($b_{\mathbf{q}}$) create (annihilate) a
phonon with wave vector $\mathbf{q}$ and linear dispersion
$\omega_{\mathbf{q}} = c_{LA}|\mathbf{q}|$, where $c_{LA}$ denotes the
longitudinal acoustic sound velocity. We take as parameters $D_e =
7\,\rm{eV}$, $D_h=-3.5\,\rm{eV}$ for a 5~nm sized QD with definitions for the
the deformation potential coupling $g_{\mathbf{q}}$ taken from
Ref.~\cite{krummheuer2005pur}.

\end{section}

\begin{section}{Chirped pulse excitation}

The exciton system is driven by a chirped Gaussian pulse. Such pulses are
typically obtained from a transform-limited original pulse
\begin{align*}
\Omega_0(t)=\frac{\Theta}{\sqrt{2\pi}\tau_{0}} \exp
\left(-\frac{t^{2}}{2\tau_{0}^{2}}\right)\exp\left(-i\omega_{0}t\right)
\end{align*}
with pulse area $\Theta$, duration $\tau_{0}$, and central frequency
$\omega_{0}$, which is passed through a Gaussian chirp filter
\cite{saleh2007fun} with the chirp coefficient $\alpha$. This results in the
chirped pulse
\begin{align*}
\Omega(t)=\tilde{\Omega}(t) \exp\left(-i\omega_{0}t-\frac{iat^2}{2}\right),
\end{align*}
with the pulse envelope (instantaneous Rabi frequency)
\begin{align*}\tilde{\Omega}(t)=\frac{\Theta}{\sqrt{2\pi\tau_{0}\tau}}
\exp\left(-\frac{t^{2}}{2\tau^{2}}\right),
\end{align*}
the chirped pulse length $\tau=(\alpha^{2}/\tau_{0}^{2}+\tau_{0}^{2})^{1/2}$
and the frequency chirp rate $a=\alpha/(\alpha^{2}+\tau_{0}^{4})$. Note that
$\tau$ is related to the full width at half maximum of the intensity by
$\tau_{\textrm{FWHM}}=2\sqrt{\ln 2} \tau$.\\

\end{section}

\begin{section}{Equations of motion}

The quantity of interest for the analysis of the optically induced temporal
dynamics is the reduced density matrix $\varrho$ of the exciton system
(either for the six-level or the three-level system). Without coupling to the
phonons the equations of motion are simply obtained from the Liouville-von
Neumann equation
\begin{align*}
\frac{d}{dt}\varrho = \frac{i}{\hbar}\left[\varrho, H \right].
\end{align*}
If carrier-phonon interaction is included, the equations of motion involve
phonon-assisted density matrix and they represent the starting point of an
infinite hierarchy of higher-order phonon-assisted density matrices. This
hierarchy is truncated using a fourth-order correlation expansion. The
equations of motion for the two-level system can be found in
Ref~\onlinecite{krugel2005the}, while the extension to the three-level system
is straightforward.

\end{section}

%\bibliography{std,qds,review_books,review_phonon,mn}